\documentclass[reprint,superscriptaddress,amsmath,amssymb,aps,preprintnumbers]{revtex4-1}
\usepackage{graphicx}
\usepackage{dcolumn}
\usepackage{bm}
\usepackage[usenames]{color}
\usepackage[normalem]{ulem}

\renewcommand{\sout}[1]{\bgroup \color{red} \ULdepth=-.5ex \ULset {#1}}
\newcommand{\comment}[1]{}
\graphicspath{{./Figs/}}
\newcommand{\average}[1]{\ensuremath{\langle#1\rangle}}

\newcommand{\VEV}[1]{\left\langle{#1}\right\rangle}

\newcommand{\TCF}{C_{12}}

\newcommand{\vk}{{\bm{k}}}

\newcommand{\omegak}{\omega_\vk}

\begin{document}
\preprint{KUNS-2823/YITP-20-89}
\title{Shear viscosity of classical Yang-Mills field}
\author{Hidefumi Matsuda}
\affiliation{Department of Physics, Faculty of Science, Kyoto University,
Kyoto 606-8502, Japan}
\author{Teiji Kunihiro}
\affiliation{Yukawa Institute for Theoretical Physics, Kyoto University,
Kyoto 606-8502, Japan}
\author{Berndt M\"uller}
\affiliation{Department of Physics, Duke University, Durham, NC 27708-0305, USA}
\author{Akira Ohnishi}
\affiliation{Yukawa Institute for Theoretical Physics, Kyoto University,
Kyoto 606-8502, Japan}
\author{Toru T. Takahashi}
\affiliation{National Institute of Technology, Gunma college,
Gunma 371-8530, Japan}

\begin{abstract}
We investigate the shear viscosity $\eta$ of the classical Yang-Mills (CYM) field on a lattice by using the Green-Kubo formula, where the shear viscosity is calculated from the time-correlation function of the energy-momentum tensor in equilibrium. Dependence of the shear viscosity $\eta(g,T)$ on the coupling $g$ and temperature $T$ is represented by a scaling function $f_\eta(g^2T)$ as $\eta(g,T)=Tf_\eta(g^2T)$ due to the scaling-invariant property of the CYM. The explicit functional form of $f_\eta(g^2T)$ is successfully determined from the calculated shear viscosity: It turns out that $\eta(g,T)$ of the CYM field is proportional to $1/g^{1.10-1.88}$ at weak coupling, which is a weaker dependence on $g$ than that in the leading-order perturbation theory but consistent with that of the "anomalous viscosity" $\eta\propto 1/g^{1.5}$ under the strong disordered field. The obtained shear viscosity is also found to be roughly consistent with that estimated through the analysis of the anisotropy of the pressure of the CYM dynamics in the expanding geometry with recourse to a hydrodynamic equation.
\end{abstract}

\maketitle

\section{Introduction}

It is widely believed that the initial stage of relativistic heavy-ion collisions is well described by the classical Yang-Mills (CYM) field~\cite{MVinitial,Glasma:instability,EG:instability}. The real-time lattice simulations of the CYM field have provided important insights into the non-equilibrium dynamics of this dense gluon matter,
which is known as {\em glasma}. In the current understanding of the early dynamics of the heavy-ion collisions, the instabilities~\cite{Glasma:instability,EG:instability,Weibel:instability,Savvidy:instability,NO:instability,Parametric:instability}
and the chaoticity~\cite{YM-chaos} of the CYM field lead to a rapid thermalization of the glasma, so that a hydrodynamic description makes sense~\cite{Heinz:RHIC}. The hydrodynamic analyses showed that the created matter may be an almost perfect fluid whose shear viscosity is close to the lower bound predicted by superstring theories with an Einsteinian classical limit~\cite{KSSbound}. The mechanism that leads such a near-minimal shear viscosity has been studied from various points of view~\cite{Arnold:shear,ABM:shear,Xu:shear,HP:shear,Homor:shear,Kyoto:scalarshear}.

Then a natural question that arises here is whether the CYM-field description of the early stage can be further extended to account for the transition region from the glasma to the low-viscosity, near-thermal gluonic fluid. To answer this question, the equilibrium transport properties of the CYM field must be investigated. It was shown in Ref.~\cite{Kyoto:entropy} that the system reaches a quasi-stationary state and the entropy increases slowly after a rapid isotropization of pressure. In Ref.~\cite{EG:instability}, the shear viscosity is deduced through the fitting of the time dependence of the energy density of the CYM field with that expected as a solution of the viscous hydrodynamics equation, while the pressure anisotropy still remains. These analyses suggest that the CYM field may be relaxed by the intrinsic dynamics to a quasi-local equilibrium state. Hence it should make sense to explore its transport properties such as the viscosities.

One of the ways to extract the shear viscosity of the CYM field is to fit the hydrodynamic parameters to the non-equilibrium time evolution of the CYM field as done in Ref.~\cite{EG:instability}. Another, more standard way is to use the Green-Kubo formula obtained in the linear response theory.
The latter approach has been successfully applied to the classical scalar theory~\cite{Homor:shear,Kyoto:scalarshear}.

In this article, we investigate the real-time dynamics of CYM field close to thermal equilibrium and extract the shear viscosity of the CYM field at thermal equilibrium. We employ the Green-Kubo formula in order to investigate the shear viscosity arising from the CYM field's dynamics. In the Green-Kubo approach, the shear viscosity can be extracted from the thermal expectation value of the time-correlation function of the energy-momentum tensor. We evaluate the expectation value as the ensemble average measured with thermally equilibrated CYM field configurations. We also make a detailed analysis of the time-correlation function and the power spectrum (the Fourier transform of the time-correlation function) in order to gain deeper insight into the relaxation of the energy-momentum tensor and the origin of the shear viscosity.

It is worth noting that our analysis is based on the scaling property of the CYM theory. The classical equation of motion of the CYM theory is invariant under the scale transformation, $g \to \gamma g$, $A \to A/\gamma$ and $E \to E/\gamma$. As the thermal classical gauge theory must be regulated on a lattice with spacing $a$, this scale invariance implies that intensive quantities in the thermal CYM theory, such as the shear viscosity, are functions of the dimensionless quantity $g^2Ta$. Setting $a=1$, any intensive quantity thus becomes a function of the product of the squared coupling $g^2$ and the temperature $T$ measured in lattice units. One of the main results of the present work thus is to determine the scaling function of the shear viscosity, $f_\eta(g^2T)$, over a wide range of the scaling variable $g^2T$.

This article is organized as follows: In Sect.~\ref{Sec:Theory}, we introduce the lattice formulation of the CYM field, the Green-Kubo formula for the shear viscosity, the classical field ensemble, and some properties of the CYM field on a lattice. In Sect.~\ref{Sec:Results}, we show the numerical results of the time-correlation function of the energy-momentum tensor, the power spectrum, and the shear viscosity. We discuss the $g$ and $T$ dependence of the obtained shear viscosity in terms of the lattice unit by using the scaling functions. In Sect.~\ref{Sec:Summary}, we summarize our work.


\section{Classical Yang-Mills Field and Its Shear Viscosity}\label{Sec:Theory}
\subsection{Classical Yang-Mills Field Theory on Lattice}

We consider the CYM theory on a $L^3$ lattice in the temporal gauge, $A^a_0(x)=0$. Its Hamiltonian in the non-compact formalism is given as (using $a=1$):
\begin{align}
H=\frac{1}{2}\sum_{\bm{x},a,i}\left(E^a_i(x)^2 + B^a_i(x)^2\right)\label{hamiltonian}.
\end{align}
Here the electric and magnetic gauge fields, $\left(E^a_i(x), B^a_i(x)\right)$, are given by
\begin{align}
E^a_i(x)&=\dot{A}^a_i(x),\\
B^a_i(x)&=\varepsilon^{ijk}F^a_{jk}(x),\\
F^a_{jk}(x)&= \partial^{\rm F}_iA^a_j(x)-\partial^{\rm F}_jA^a_i(x)\\\nonumber 
&-\!\frac{g^2}{4}\varepsilon^{abc}\!
\left[A^b_i(x)+A^b_i(x\!+\!\hat{j})\right]\left[A^c_j(x)+A^c_j(x\!+\!\hat{i})\right],
\end{align}
where $(A^a_i(x), E^a_i(x))$ are the canonical variables and $\partial^{\rm F}_i$ is the forward difference operator. The time evolution of the CYM field is obtained by solving the Hamilton equation of motion,
\begin{align}
\dot{A}^a_i(x) = \frac{\partial H}{\partial E^a_i(x)}\ ,\quad
\dot{E}^a_i(x) = -\frac{\partial H}{\partial A^a_i(x)}.
\label{Eq:EOM}
\end{align}

By solving this equation of motion, we calculate the time-evolution of the space-averaged off-diagonal matrix elements of the energy-momentum tensor defined as
\begin{align}
\tau_{ij}(t)\equiv\frac{1}{L^3}\sum_{\bm{x,a}}\left[ E'^a_i(x)E'^a_j(x) + B'^a_i(x)B'^a_j(x)\right],\label{emtensor}
\end{align}
on the lattice. Here $(E'^a_i(x),B'^a_i(x))$ are the electric and magnetic fields defined at a shifted spatial point
\begin{align}
&E'^a_i(x)\equiv\left( E^a_i(x) + E^a_i(x-\hat{i}) \right)/2,\\
&B'^a_i(x)\equiv\left( B^a_i(x) + B^a_i(x+\hat{i}) \right)/2.
\end{align}
It should be noted that the electric and magnetic fields $(E'^a_i(x),B'^a_i(x))$ are defined at the same spatial position for different $i$ because $A^a_i(x)$ resides on the link between $x$ and $x+\hat{i}$.  Therefore, $(E'^a_i(x),B'^a_i(x))$ make the matrix elements more symmetric in the space directions
than $(E^a_i(x),B^a_i(x))$.


\subsection{Green-Kubo Relation}

The shear viscosity in any systems close to equilibrium can be evaluated by means of the Green-Kubo formula~\cite{Zubarev:statistical}:
\begin{align}
\eta =& 
\frac{1}{T} \int_0^\infty dt\, C(t),
\label{Eq:GK}
\\
C(t) =& \frac{1}{3}
\sum_{i<j}
V \average{\tau_{ij}(t)\tau_{ij}(0)},
\label{Eq:C12}
\end{align}
where $T$ and $V$ denote temperature and volume in lattice units, respectively. $\tau_{ij}(t)$ is the space-averaged off-diagonal matrix element of the energy-momentum tensor, and $\langle \cdots\rangle$ denotes the expectation value in equilibrium. $C(t)$ is the direction-averaged time-correlation function of $\tau_{ij}(t)$.

While the correlation disappears in the long time-separation limit $t\rightarrow \infty$, and thus $C(t)$ is expected to approach $V\,\average{\tau_{ij}(0)}^2_{\rm eq}=0$ in this limit, the integral in \eqref{Eq:GK} converges only when the correlation function falls off faster than $1/t$ or exhibits oscillatory behavior with a decreasing amplitude. We will check the convergence in Sect.~\ref{Sec:fs}. We will also compute the Fourier transform of $C(|t|)$,
\begin{align}
\rho(\omega)=\int_0^\infty dt\,\cos\omega t\,C(t)
=\frac{1}{2}\int^\infty_{-\infty} dt e^{i\omega t} C(|t|) ,
\end{align}
and discuss the structure of $\rho(\omega)$ in comparison with that in the scalar theory in Appendix \ref{App:strong}.


\subsection{Classical Field Ensemble}\label{Sec:ensemble}

For the calculation of the shear viscosity via the Green-Kubo formula, we need the thermal expectation value of the time correlation of the energy-momentum tensor. We evaluate the expectation value from the ensemble average of the classical field configurations in thermal equilibrium as
\begin{align}
\average{ \mathcal{O} }
\simeq& \frac{1}{N_\mathrm{conf}}\sum^{{N_\mathrm{conf}}}_{i=1} \mathcal{O}_i,
\label{ensemble}
\end{align}
where $N_\mathrm{conf}$ is the total number of configurations used in the evaluation, and $\mathcal{O}_i$ is the observable measured with the $i$-th configuration. Statistical errors of the expectation values, which are of the order of $1/\sqrt{N_\mathrm{conf}}$, are estimated and taken into account. We prepare the ensemble of thermally equilibrated CYM field configurations
by the long time evolution of the classical CYM field starting from the random initial configurations. The procedure we have used is explained in detail in Sect.~\ref{Sec:Set}.


\subsection{Equipartition of Electric Field Energy}\label{Sec:equi}

The electric field strength can be utilized to examine thermalization and also to measure the temperature. In equilibrium, the distribution of canonical variables $(A,E)$ is given by
the canonical partition function $\mathcal{Z}$,
\begin{align}
\mathcal{Z}
=& \int \mathcal{D}E \mathcal{D}A \exp(-H/T)
\nonumber\\
=&\mathcal{Z}_B[A] \times\prod_{\bm{x},a,i}
\int dE^a_i(\bm{x}) e^{-(E^a_i(\bm{x}))^2/2T}
\nonumber\\
=&\mathcal{Z}_B[A] \times\prod_{\bm{k},a,i}
\int dE^a_i(\bm{k}) e^{-|E^a_i(\bm{k})|^2/2T}
\ ,\label{partition}
\end{align}
where $\mathcal{Z}_B=\int \mathcal{D}A\,\exp(-H_B/T)$ is the magnetic field part of the partition function:
\begin{align}
H_B=\sum_{\bm{x},a,i}(B_i^a(x))^2/2.
\end{align}
From the second line in \eqref{partition}, the average squared electric field is found to be equal to the temperature, $\average{(E^a_i(\bm{x}))^2}=T$ at each spatial position $\bm{x}$. From the third line, we find that the same relation
\begin{align}
\average{|E^a_i(\bm{k})|^2}=T
\label{Eq:equi} ,
\end{align}
applies to the mean square Fourier component of the electric field
\begin{align}
E^a_i(\bm{k})=\sum_{\bm{x}}E^a_i(\bm{x})e^{-i\bm{k}\cdot\bm{x}}/\sqrt{L^3} ,
\end{align}
for each lattice momentum $\bm{k}=2\pi(n_x,n_y,n_z)/L$ ($n_i=0, 1, \cdots L-1$). Note that $E^a_i(\bm{k})$ and $E^a_i(-\bm{k})$ are not independent but related as $E^a_i(-\bm{k})=(E^a_i(\bm{k}))^*$, since $E^a_i(\bm{x})$ is real. Note that, while $E^a_i(\bm{k})$ itself is not gauge invariant, the property \eqref{Eq:equi} does not depend on the chosen gauge. In Sect.~\ref{Sec:Set}, we will examine thermalization from the momentum independence of $\average{|E^a_i(\bm{k})|^2}$ and obtain the temperature from $\average{|E^a_i(\bm{k})|^2}$.


\subsection{Scaling Property}

Here we explain the scaling property of the CYM theory, which plays an important role in our analysis of the shear viscosity.
The equation of motion of the CYM field is invariant under the following scaling transformation,
\begin{align}
A^a_i(x) \to A^a_i(x)/\gamma,\ 
E^a_i(x) \to E^a_i(x)/\gamma,\
g \to \gamma g.
\label{Eq:scaling}
\end{align}
Then, the lattice temperature defined in Eq.~\eqref{Eq:equi} scales as
\begin{align}
T_g
&=\average{|E(\bm{k})|^2}_g
=\gamma^2 \average{|E(\bm{k})/\gamma|^2}_g\nonumber\\
&=\gamma^2 \average{|E(\bm{k})|^2}_{\gamma g}
=\gamma^2 T_{\gamma g},
\end{align}
where we have omitted the color and vector indices. This relation can be rewritten as $(\gamma g)^2 T_{\gamma g}=g^2T_g$. Thus, once we prepare the ensemble at a given value of $g^2T$, we can obtain the thermal average of observables at various values of $(g,T)$ having the same $g^2T$ by simply rescaling the observables. For example, the time-correlation function $C(t)$, its power spectrum (the Fourier transform of $C(t)$) $\rho(\omega)$, and the shear viscosity $\eta$ are given as
\begin{align}
C(t;g,T)
= &\gamma^4C(t;\gamma g,T/\gamma^2),
\\
\rho(\omega;g,T)
= &\gamma^4\rho(\omega;\gamma g,T/\gamma^2),
\\
\eta(g,T)
= &\gamma^2\eta(\gamma g,T/\gamma^2), 
\label{scaling}
\end{align}
where we have used the scaling property of the energy-momentum tensor,
\begin{align}
T_{\mu\nu}(A'=A/\gamma,E'=E/\gamma)=T_{\mu\nu}(A,E)/\gamma^2.
\end{align}
These relations show that there exist scaling functions of $g^2T$
\begin{align}
f_C(t,g^2T)=&C(t;g,T)/T^2,\\ 
f_\rho(\omega,g^2T)=&\rho(\omega,g,T)/T^2,\\
f_\eta(g^2T)=&\eta(g,T)/T.
\label{Eq:scalingfunc}
\end{align}
These scaling functions are related with each other as follows,
\begin{align}
f_\eta(g^2T) &= \lim_{\omega \to 0}\int^\infty_0 dt e^{i\omega t} f_C(t,g^2T).\\
f_\rho(\omega,g^2T) &= \int^\infty_0 dt \cos{\omega t} ~ f_C(t,g^2T),
\end{align}
When multiplied by appropriate powers of $T$, these scaling functions give the values of $C(t;g,T)$, $\rho(\omega;g,T)$, and $\eta(g,T)$. In this study, we perform calculations at several values of $g$ with $T \approx 1$. Results will shown in terms of the above scaling functions.


\section{Numerical results}
\label{Sec:Results}


\subsection{Calculation Setup and Initial Configuration}\label{Sec:Set}

Our calculations were performed on $16^3$, $24^3$ and $32^3$ lattices with periodic boundary conditions for the SU(2) Yang-Mills field keeping the average energy per degree of freedom at unity, which corresponds to the CYM system with the temperature of $T \approx 1$. The coupling constant was varied in the wide range $g=0.15-20$. The equation of motion is solved by leapfrog integration with $\Delta t=0.01$. The thermal expectation value of an observable is evaluated from the ensemble average using  $N_\mathrm{conf}=1000$ thermal configurations.

\begin{figure}[htbp]
\begin{center}
\includegraphics[width=0.99\linewidth]{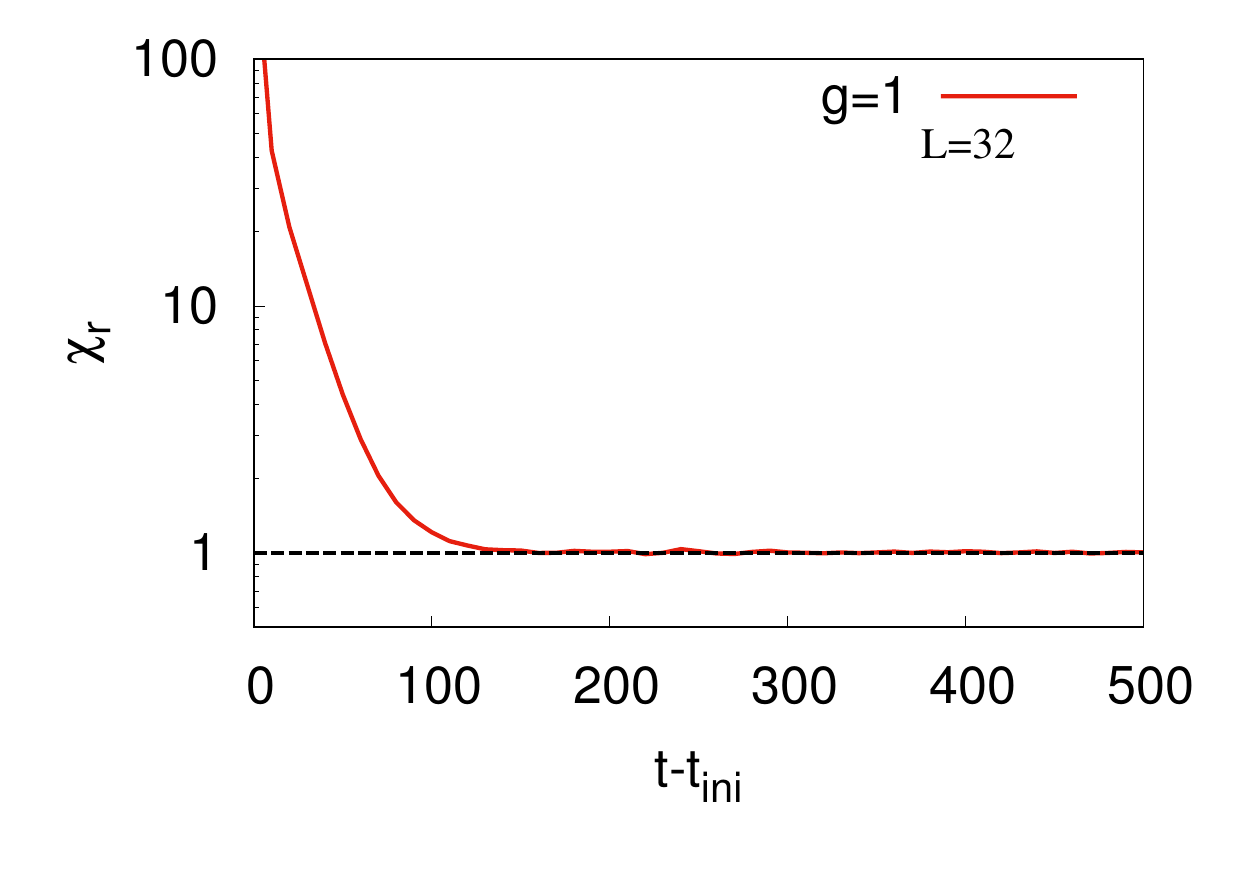}
\end{center}
\caption{The time evolution of the reduced chi-squared from the fit to the ensemble average, $\VEV{|E^1_1(\bm{k})|^2}=T_\mathrm{fit}$,
at $g=1$ on the $32^3$ lattice.
}
\label{Fig:chi}
\end{figure}

Here we explain the procedure for generating a set of thermally equilibrated gauge field configurations. First, we generate totally $N_{\rm conf}$ field configurations, $(A^{(i)},E^{(i)})\ (i=1,2,\cdots ,N_{\rm conf})$, at initial time $t=t_{\rm ini}$, as
\begin{align}
A^{(i),a}_j(t=t_{\rm ini},\bm{x}) &= 0\ \ \ (j=1,2,3),\\
E^{(i),a}_\perp(t=t_{\rm ini},\bm{x})&=R^{(i),a}_\perp(\bm{x})\partial^B_3 R^{(i),a}_3(\bm{x})/N_E,\label{ini}, \\
E^{(i),a}_3(t=t_{\rm ini},\bm{x}) &=-R^{(i),a}_3(\bm{x})\partial^B_\perp \cdot R^{(i),a}_\perp(\bm{x})/N_E,
\end{align}
where
$\partial^{\rm B}_i$ is the backward difference operator, $R^{(i),a}_j(\bm{x})$ is a Gaussian random number with unit variance:
\begin{align}
\VEV{R^{(i),a}_j(\bm{x})R^{(i'),a'}_{j'}(\bm{y})} &= \delta_{ii'}\delta_{aa'}\delta_{jj'}\delta_{\bm{x}\bm{y}},
\end{align}
and $N_E$ is a normalization factor chosen to make the average energy per each degree of freedom equal to unity. This initial configuration satisfies the lattice Gauss law:
\begin{align}
&\bm{\partial^{\rm B}}\!\cdot\!\bm{E}^a(x)\!%
-\!\frac{f^{abc}}{2} \left[\bm{A}^b(x)\!\cdot\!\bm{E}^c(x)\right.
\nonumber\\
&+ \left. \bm{A}^b(x\!-\!\hat{i})\!\cdot\!\bm{E}^c(x\!-\!\hat{i}) \right] = 0.
\end{align}
Next, we evolve each configuration with the classical equation of motion until it equilibrates. We confirm the equilibration by checking for equipartition of the energy given in Eq.~\eqref{Eq:equi}. In Fig.~\ref{Fig:chi}, we show the time evolution of the reduced chi-square of the fit $\VEV{|E^1_1(\bm{k})|^2}=T_\mathrm{fit}$ with $T_\mathrm{fit}$ as fitting parameter using the $N_{\rm conf}=1000$ configurations with $g=1$ and $L=32$. The reduced chi-squared value reaches unity for $t-t_{\rm ini}>t_{eq}\sim200$, which indicates that Eq.~\eqref{Eq:equi} holds after this time. We can consider the configurations at $t-t_{\rm ini}>t_{eq}\sim200$ to be in thermal equilibrium for $t-t_{\rm ini}>t_{eq}$.
In Appendix~\ref{App:local}, we confirm the equilibration using another quantity, the local electric energy distribution.

\begin{figure}[tbhp]
\begin{center}
\includegraphics[width=0.99\linewidth]{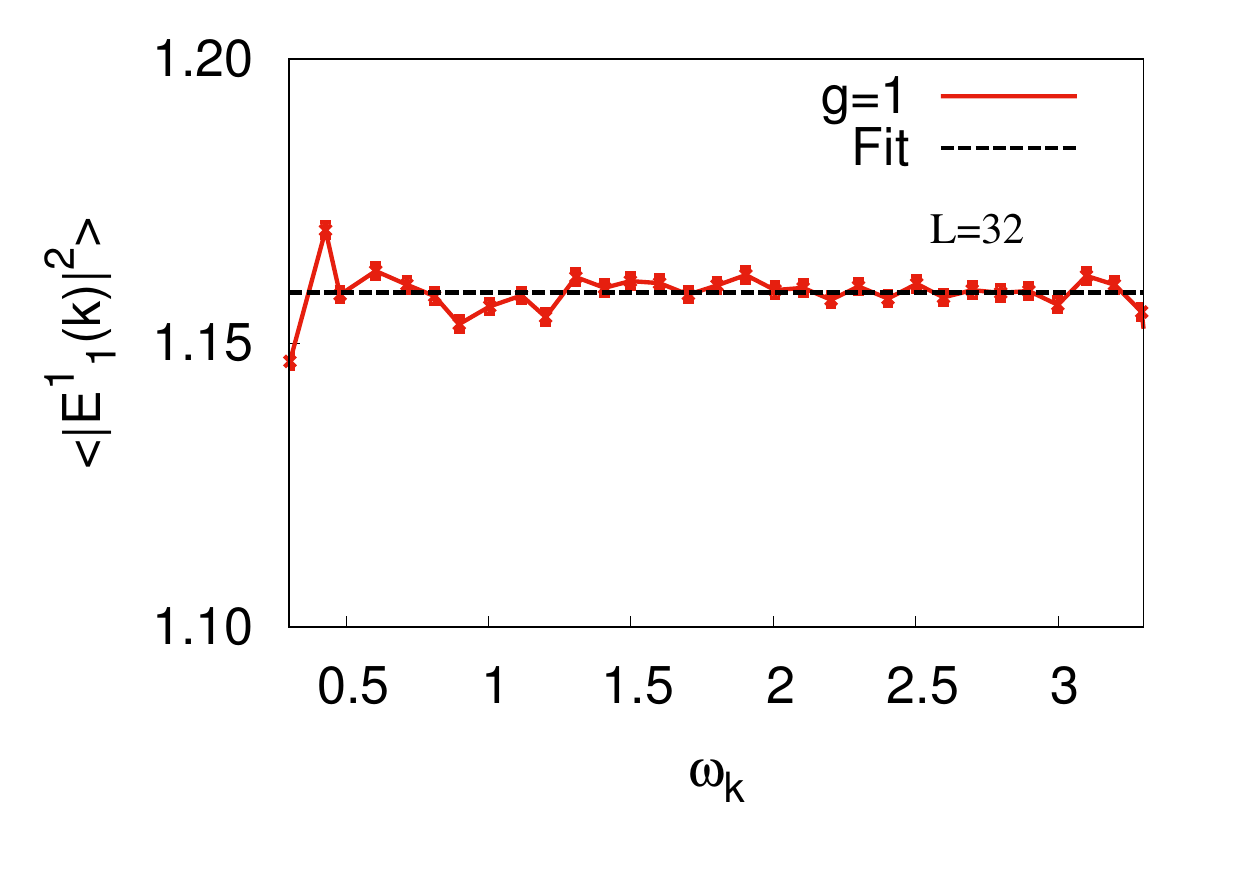}
\end{center}
\caption{Thermal expectation value of $|E^1_1(\bm{k})|^2$ as a function of frequency, $\omegak$, at $g=1$ on the $L=32^3$ lattice.
Black dashed line shows
the constant fit function, $T_\mathrm{fit}$}.
\label{figEP}
\end{figure}

Finally, we extract the temperature of the thus prepared configurations via Eq.~\eqref{Eq:equi}. In Fig.~\ref{figEP}, we show $\average{|E_1^1(\bm{k})|^2}$ as a function of the lattice frequency $\omegak$,
\begin{align}
\omegak = 2 \sqrt{\sin^2(k_1/2)+\sin^2(k_2/2)+\sin^2(k_3/2)},
\label{Eq:omegak}
\end{align}
at $g=1$ on the $32^3$ lattice. The expectation value has been evaluated from the configurations at $t-t_{\rm ini}=200$. It appears that $\average{|E_1^1(\bm{k})|^2}$ is around unity at
any $\omegak$ as we expected. By fitting a constant $T_\mathrm{fit}$ to the result, we obtain
\begin{eqnarray}
T_{\rm fit} = 1.16\ \ (\chi^2_{\rm r}=1.00).
\end{eqnarray}
The reduced chi-squared is around unity ($\chi^2_{\rm r}\sim1$), and thus we can consider the fitted value of $T_{\rm fit} = 1.16$ as the temperature of the system with $(g, L)=(1,32)$.
In Table~\ref{Tab:eta} further below, we summarize the values of $g^2T$ used in our calculations; they cover the range $0.0256 \leq g^2T \leq 499$.

\subsection{Time correlation function of energy-momentum tensor}
\label{Sec:TCF}

We shall now discuss the time-correlation function of the energy-momentum tensor, $C(t)$, evaluated in the thermally equilibrated CYM field.

\begin{figure}[tbhp]
\begin{center}
\includegraphics[width=0.99\linewidth]{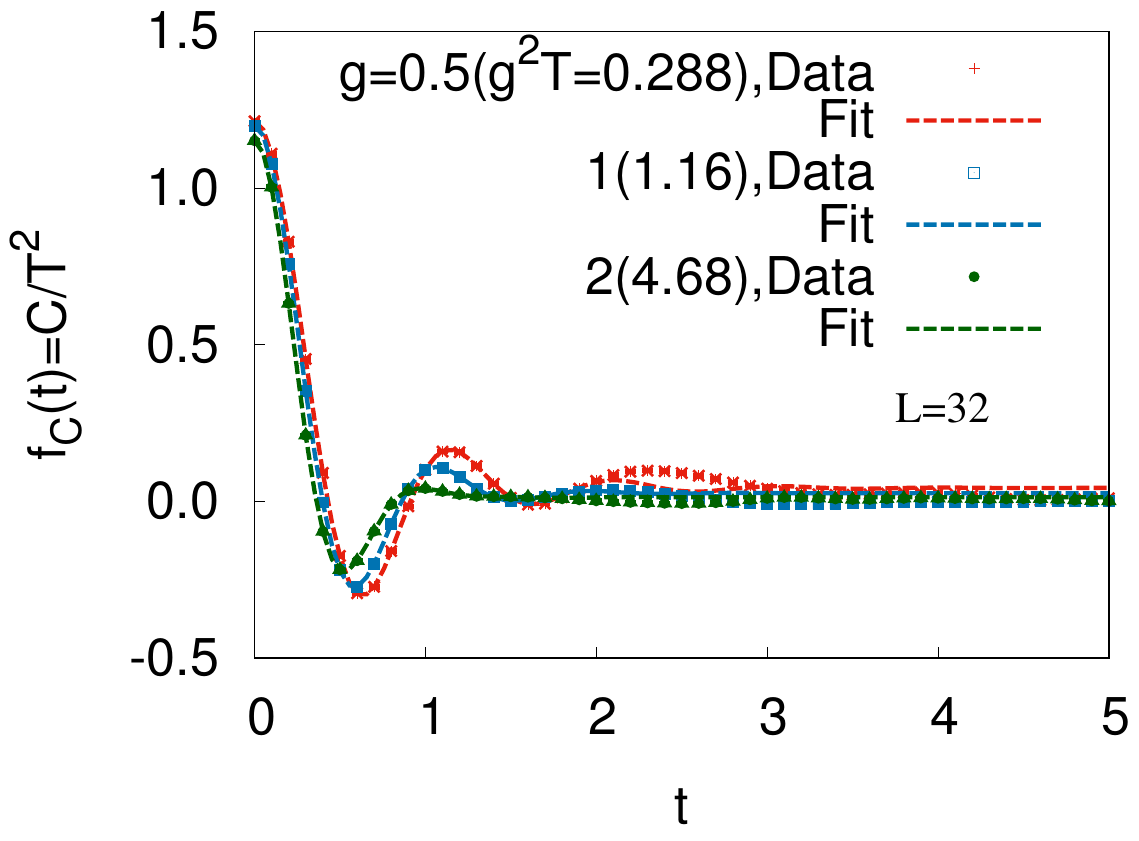}
\end{center}
\caption{
Time correlation functions of the energy-momentum tensor normalized by $T^2$,
$f_C(t)=\sum_{i<j}V\average{\tau_{ij}(t)\tau_{ij}(0)}/3T^2$ in the short time range($t<5$). 
We show the results at $g=0.5, 1.0$ and $1.5$ ($g^2=0.288, 1.16$ and $4.68$) 
on the $32^3$ lattice by symbols.
Dashed lines show the fitting results in the form of
$F_{\rm DO}(x)(=\sum_{i=1,2} a_ie^{-b_ix}\cos{(c_ix+d_i)})+{\rm const}$.
}
\label{Fig:TCFshort}
\end{figure}

In Fig.~\ref{Fig:TCFshort}, we show the normalized time-correlation functions, $f_C(t)=C(t)/T^2$, in the short time range, $t<5$. The time-correlation function is seen to decrease rapidly over the interval $0<t<1.5$. In this range it can be well described by the sum of two damped oscillators ,
\begin{align}
F_{\rm DO}(x)= \sum_{i=1,2} a_ie^{-b_ix}\cos{(c_ix+d_i)},
\end{align}
superimposed on a constant, as shown by the dashed lines. Therefore, the rapid decrease in the short time range is found to come from the two damped oscillatory motions.

\begin{figure}[tbhp]
\begin{center}
\includegraphics[width=0.99\linewidth]{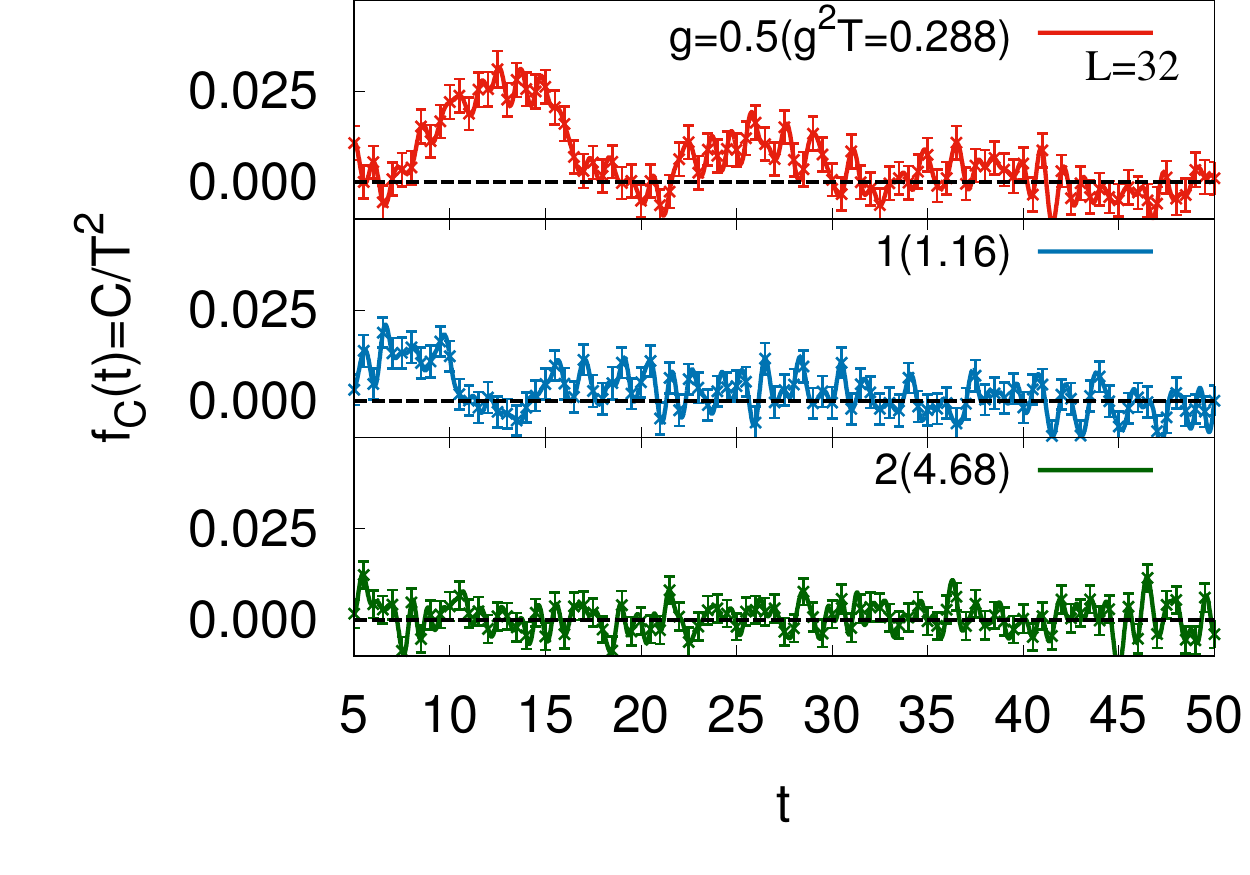}
\end{center}
\caption{The same as Fig.~\ref{Fig:TCFshort} but in the long-time range,}
\label{Fig:TCFlong}
\end{figure}

In Fig.~\ref{Fig:TCFlong}, we show $f_C(t)$ in the long-time range, $5<t<50$. The remaining correlation is found to decay more slowly in this later time period, and the decay time scale significantly depends on the value of $g^2T$.

\subsection{Power spectrum}\label{Sec:fs}

Before discussing the shear viscosity, we consider the Fourier transform of the time-correlation function $C(|t|)$, referred to as the power spectrum, whose low-frequency limit gives the shear viscosity. The power spectrum of $C(|t|)$ is given by
\begin{align}
&\rho(\omega)
=\int^\infty_0 dt\, \cos\omega t\, C(t)
\nonumber\\
&=\frac12\lim_{t_\mathrm{cut}\rightarrow\infty}
\int^{t_\mathrm{cut}}_{-t_\mathrm{cut}} dt e^{i\omega t} C(|t|)
\nonumber\\
&\sim \lim_{\Delta t\to0, N_t \to \infty}\frac{\Delta t}{2} \sum_{n=-N_t}^{N_t-1} e^{i\omega n \Delta t} C(|n \Delta t|)
,\label{Eq:Fourier}
\end{align}
where $\Delta t$ is the time step size and $N_t$ is the total number of steps in the time evolution, and is half of the data points in summation. In the last line, we represent the integration over the interval of $[-\infty,\infty]$ by the discrete Fourier transformation with the period $2N_t$, in which the power spectrum $\rho(\omega)$ has entries at each discrete frequency $\omega=2\pi n/(2N_t \Delta t)$ ($n=-N_t,-N_t+1,\cdots N_t-1$).

\begin{figure}[tbhp]
\begin{center}
\includegraphics[width=0.99\linewidth]{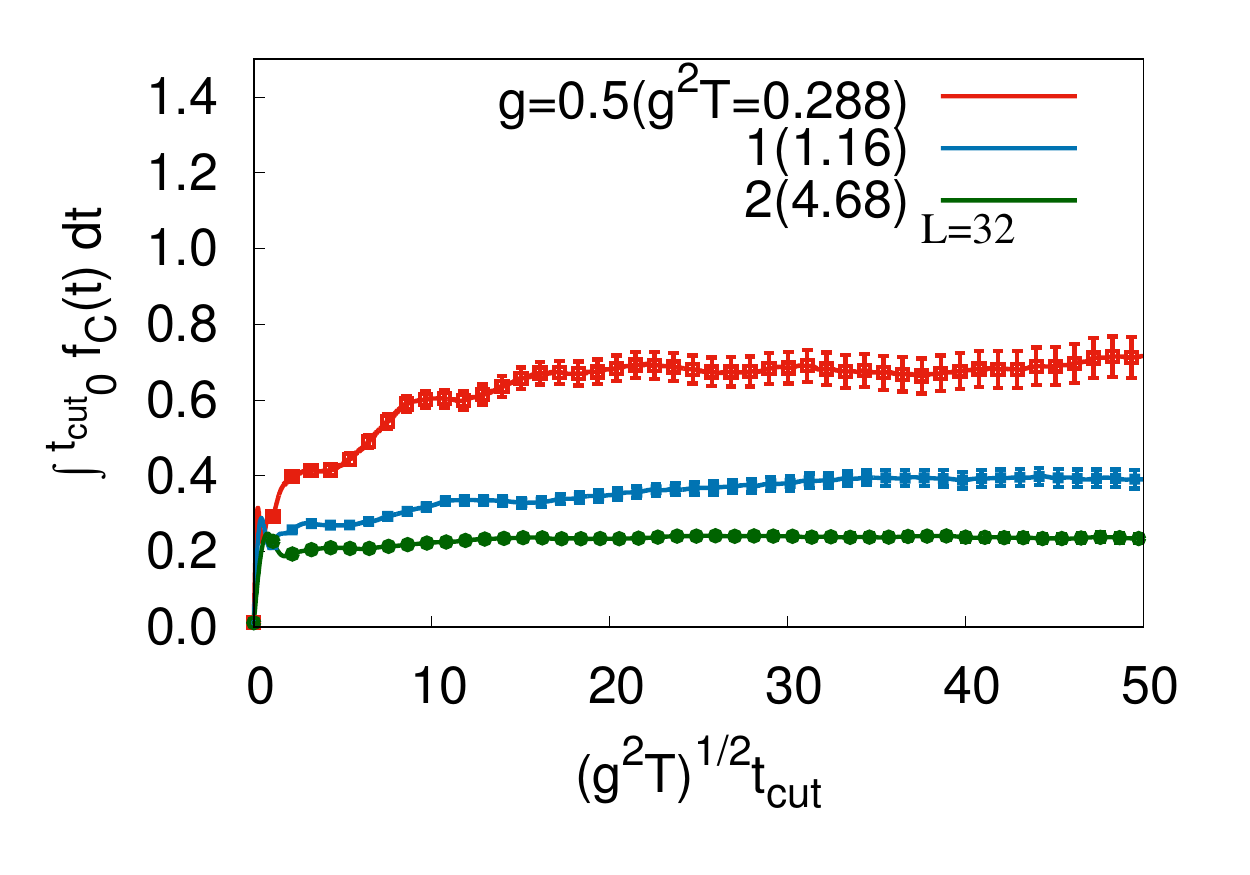}
\end{center}
\caption{Convergence of the integral of the time-correlation function normalized by $T^2$, $\int_0^{t_\mathrm{cut}} dt\,f_C(t)=\int_0^{t_\mathrm{cut}} dt\,C(t)/T^2$ on the $32^3$ lattice at $g=0.5, 1.0, 2.0$ ($g^2T=0.288, 1.16, 4.68$).
}
\label{Fig:conv}
\end{figure}

In actual calculations, the upper bound of the integration, $t_\mathrm{cut}=N_t \Delta t$, is finite but must be large enough so that the integral (summation) in \eqref{Eq:Fourier} converges. As an example of the convergence check, we show in Fig.~\ref{Fig:conv} the integral of the normalized time-correlation function $C(t)/T^2$ over the interval $0 \leq t \leq t_\mathrm{cut} = N_t\Delta t$, which is directly related to the shear viscosity $\eta$ via \eqref{Eq:GK} and corresponds to $\rho(0)$ in \eqref{Eq:Fourier}. The integral at each value of $g^2T$ shows convergent behavior at large $t_\mathrm{cut}$, while small oscillations around a constant are still visible. Thus, the integral in Eq.\eqref{Eq:Fourier} can be approximated by the integral over the finite interval
as long as $N_t$ is large enough. This is in contrast with the classical $\phi^4$ theory case, in which a very long tail appears in the time-correlation function and must be subtracted in advance~\cite{Kyoto:scalarshear}.

\begin{figure}[tbhp]
\begin{center}
\includegraphics[width=0.99\linewidth]{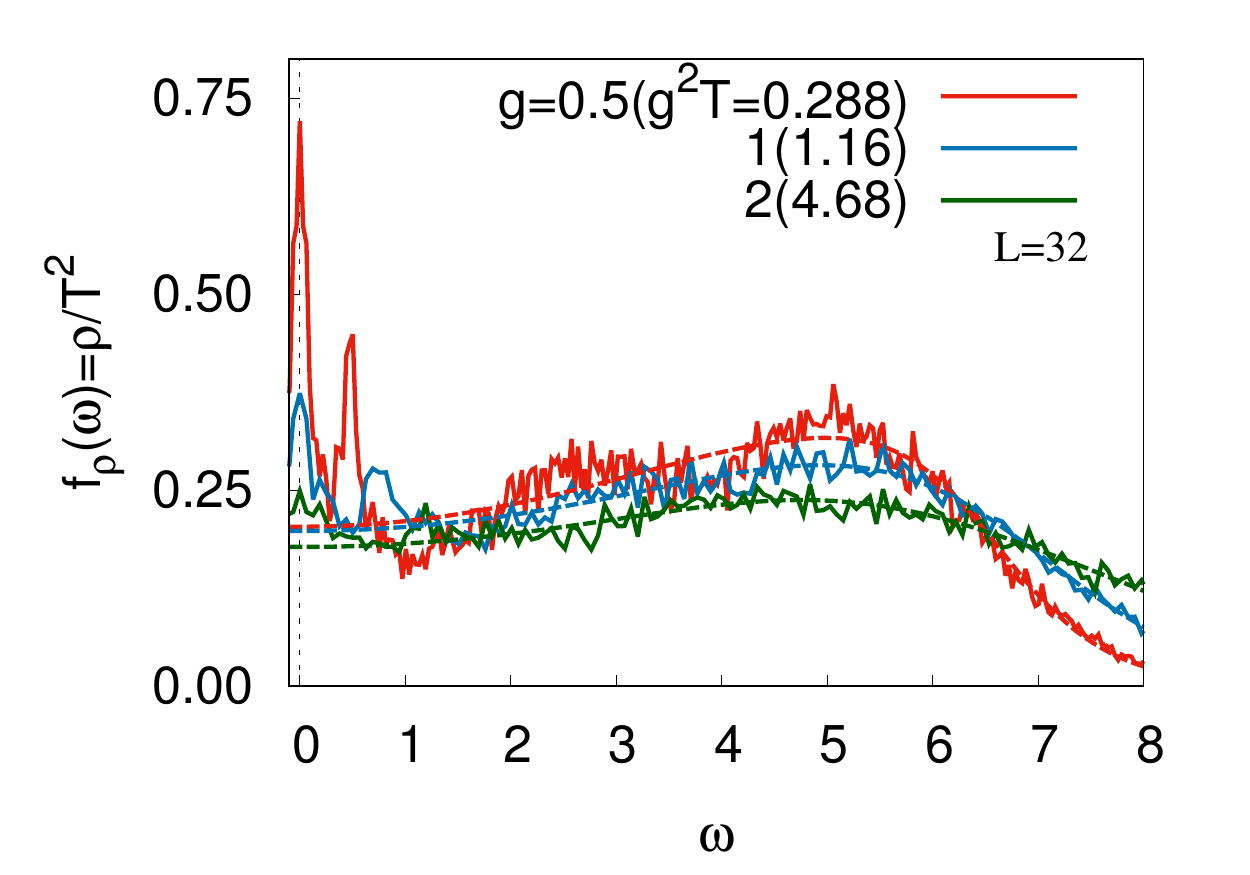}
\end{center}
\caption{Power spectra normalized by $T^2$, $f_\rho(\omega)=\rho(\omega)/T^2$, at $g=0.5, 1.0, 1.5$ ($g^2=0.288, 1.16, 4.68$) 
on the $32^3$ lattice shown by solid curves. The dashed curves depict the Fourier transform of the damped oscillatory part of the fitting function in Fig.~\ref{Fig:TCFshort}, $F_{\rm DO}(|t|)$.}
\label{Fig:Spec}
\end{figure}

In Fig.~\ref{Fig:Spec}, we show the normalized Fourier spectra, $f_\rho(\omega)=\rho(\omega)/T^2$, at $g=0.5, 1.0, 2.0$ ($g^2T=0.288, 1.16, 4.68$)
as solid curves. Dashed curves show the Fourier transform of the damped oscillator part of the fitting function, $F_{\rm DO}(|t|)$, shown in Fig.~\ref{Fig:TCFshort}. The damped oscillator part corresponds to the broad bump of the spectral function, which has large strength in the high frequency region, $4 \leq \omega \leq 6$. The remaining part of the power spectrum corresponds to the long-time decay part of the time-correlation function and show the peaks at $\omega \leq 1$.
The number of peaks decreases and the peak heights become smaller with increasing $g^2T$.
The peak at $\omega=0$ remains even at large $g^2T$, and is the highest peak in the $g^2T$ region shown here.
The $\omega=0$ peak is sharp at small $g^2T$ but seems to be a continuous function of $\omega$. This supports the convergence of the integral in \eqref{Eq:Fourier} at $\omega \to 0$.
It would be interesting to discuss origin of the peaks other that at $\omega=0$, which may be due collective modes in the low frequency region. However this is beyond the scope of this work, which is concerned with the shear viscosity. At larger values of $g^2T$, the power spectrum becomes smoother and is increasingly dominated by the damped oscillator part. In Appendix~\ref{App:strong}, we show $f_\rho(\omega)$ in the extremely large $g^2T$ region.

The broad bump in the high frequency region caused by the damped oscillatory behavior in the time-correlation function was also found in Ref.~\cite{Kyoto:scalarshear}, where it was conjectured that the damped oscillatory behavior is caused by the propagation of two modes having momenta of $\bm{k}$ and $-\bm{k}$ (two-momentum mode),
\begin{align}
\TCF(t)\sim \frac{2}{V}\sum_{\bm{k}}
k_x^2k_y^2
\VEV{\phi_{\bm{k}}(t)\phi^*_{\bm{k}}(0)}
\VEV{\phi_{\bm{-k}}(t)\phi^*_{\bm{-k}}(0)}\ ,
\end{align}
where $\phi_{\bm{k}}$ denotes the Fourier transform of the classical scalar field and $\VEV{\cdots}$ denotes the thermal expectation value. The frequency difference between different two-momentum modes causes the phase decoherence and the damping of the time-correlation function. Since this can take place even in the free field case and the damped oscillations in the CYM field have shapes similar to those in the $\phi^4$ theory,
the damped oscillation in the early stage and the bump around $\omega\sim 5$ may be also caused by the interference of the two-momentum modes.


\subsection{Shear viscosity}
\label{Subsec:Shear}

We now discuss the shear viscosity. The shear viscosity $\eta$ is obtained by taking the low frequency limit of the power spectrum, 
$\eta=\lim_{\omega \to 0}\rho(\omega)/T$. In Table~\ref{Tab:eta}, we summarize the $g^2T$ dependence of the normalized shear viscosity, $f_\eta(g^2 T)=\eta/T$ on the $16^3, 24^3$ and $32^3$ lattices. As shown in \eqref{Eq:scalingfunc}, $\eta/T$ is invariant under the scaling transformation given in \eqref{Eq:scaling} and a function of of the single parameter $g^2T$.

\begin{center}
\begin{table}[htbp]
\begin{tabular}{|ccc|ccc|}
\hline
$g$&$T$& $g^2T$&$\eta/T$&$\eta/T$&$\eta/T$\\
&&&$(L=16) $&$(L=24) $&$(L=32) $\\
\hline\hline
0.15&1.14&0.0256&2.28 $\pm$0.22 &1.75 $\pm$0.22 &1.89 $\pm$0.23 \\ 
0.2 &1.14&0.0457&1.50 $\pm$0.15 &1.46 $\pm$0.16 &1.44 $\pm$0.16 \\ 
0.25&1.14&0.0715&1.39 $\pm$0.13 &1.33 $\pm$0.12 &1.22 $\pm$0.12 \\ 
0.5 &1.15&0.288 &0.670$\pm$0.055&0.663$\pm$0.056&0.728$\pm$0.057\\ 
0.75&1.15&0.649 &0.549$\pm$0.034&0.436$\pm$0.034&0.467$\pm$0.034\\ 
1   &1,16&1.16  &0.369$\pm$0.025&0.347$\pm$0.026&0.380$\pm$0.025\\ 
1.5 &1.16&2.62  &0.303$\pm$0.020&0.310$\pm$0.022&0.316$\pm$0.022\\
2   &1.17&4.68  &0.349$\pm$0.020&0.257$\pm$0.019&0.254$\pm$0.018\\ 
3   &1.18&10.6  &0.189$\pm$0.016&0.228$\pm$0.017&0.238$\pm$0.016\\ 
4   &1.18&18.9  &0.203$\pm$0.016&0.197$\pm$0.016&0.212$\pm$0.015\\
5   &1.19&29.7  &0.150$\pm$0.015&0.150$\pm$0.014&0.201$\pm$0.014\\ 
10  &1.22&122   &0.157$\pm$0.012&0.134$\pm$0.011&0.171$\pm$0.012\\ 
15  &1.24&278   &0.116$\pm$0.008&0.130$\pm$0.008&0.119$\pm$0.008\\ 
20  &1.25&499   &0.123$\pm$0.006&0.116$\pm$0.006&0.105$\pm$0.006\\
\hline\hline
\end{tabular}
\caption{Shear viscosity obtained at $g^2T=0.0256-499$ on the $16^3, 24^3$ and $32^3$ lattices.
In actual calculations, we choose the coupling constant in the range of $g=0.15-20$ and set the average energy per degree of freedom as unity, which corresponds to $T \approx (1.14-1.25)$.}
\label{Tab:eta}
\end{table}
\end{center}

\begin{figure}[htbp]
\begin{center}
\includegraphics[width=0.99\linewidth]{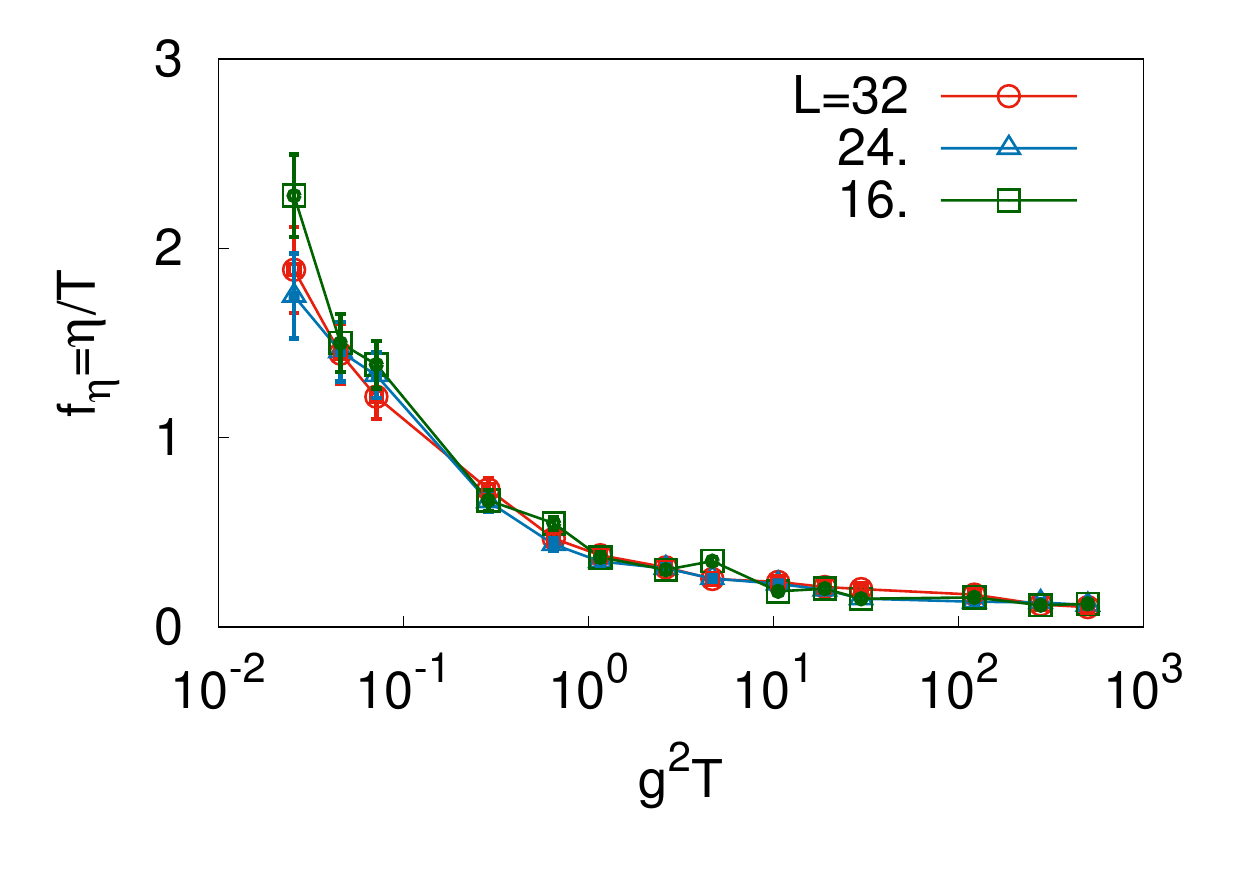}
\end{center}
\caption{
$g^2T$ dependence of the shear viscosity normalized by $1/T$, $f_\eta=\eta/T$, on the $16^3, 24^3$ and $32^3$ lattices.
}
\label{Fig:shear_scaling}
\end{figure}

In Fig.~\ref{Fig:shear_scaling}, we show $f_\eta$ as a function of $g^2T$. The results on the $16^3, 24^3$ and $32^3$ lattices are consistent with each other. We find that $f_\eta(g^2T)$ is a rapidly falling function at small values of $g^2$, but flattens out in the region $g^2T > 1$. This weaker fall-off is readily understood from the behavior of $f_C(t)$ and $f_\rho(\omega)$. The damped oscillation in $f_C(t)$ is insensitive to $g^2T$ as seen in Fig.~\ref{Fig:TCFshort}, and its contribution to $f_\rho(\omega)$ in the low frequency region, which is responsible for the viscosity, becomes more dominant and remains even in the larger $g^2T$ region.

\begin{figure}[htbp]
\begin{center}
\includegraphics[width=0.99\linewidth]{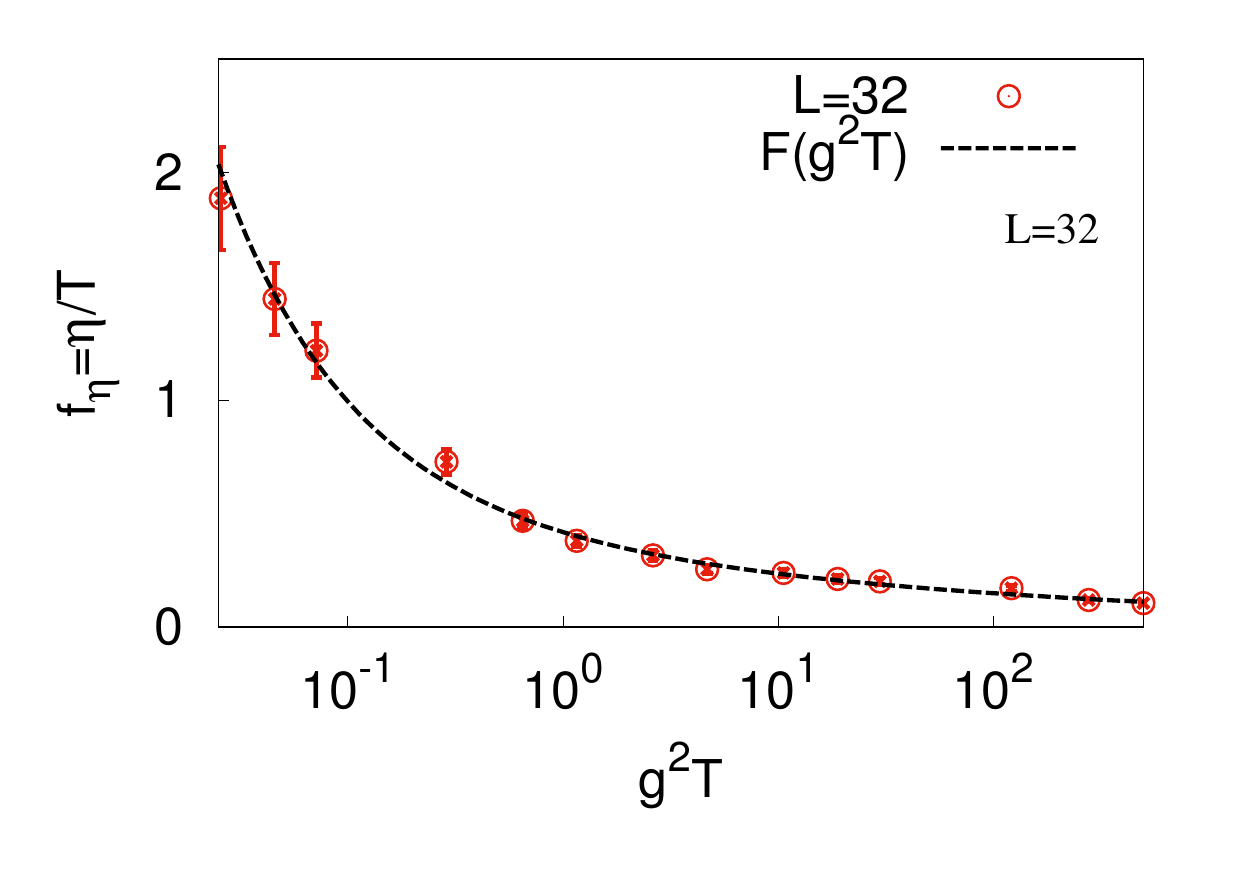}
\end{center}
\caption{
Analytical fit to the shear viscosity normalized by $1/T$, $f_\eta=\eta/T$, on the $32^3$ lattice with the fit function $F(x)=\alpha x^{-\beta/2}+\gamma x^{-\delta/2}\ (\beta>\delta)$.
}
\label{Fig:shear_fit}
\end{figure}

Next we analyze the functional form of $f_\eta(g^2T)$ and understand the $g^2T$ dependence of $f_\eta(g^2T)$ in the range, $0.0256<g^2T<499$, by fitting our results to a polynomial function with parameters $\alpha$, $\beta$, $\gamma$ and $\delta$,
\begin{align}
F(x)=\alpha x^{-\beta/2}+\gamma x^{-\delta/2} \\ 
(\beta>\delta),
\end{align}
where $x=g^2T$. The form of $F(x)$ is motivated by the results shown in Fig.~\ref{Fig:shear_scaling}. We expect that the first term of $F(x)$ describes the contribution from the long-time tail in $f_C(t)$, which seems to increase like an inverse power of $g^2T$ and dominates at small $g^2T$. The second term of $F(x)$ is expected to describe the contribution from the damped oscillation in $f_C(t)$, which is less sensitive to $g^2T$.
The fit to the $f_\eta(g^2T)$ obtained from the Green-Kubo formula in Eq.~\eqref{Eq:GK} on the $32^3$ lattice yields the following parameters:
\begin{align}
F(x):\ 
&\alpha=0.09 \pm 0.07,\  \beta=1.49 \pm 0.39,
\nonumber\\
&\gamma=0.33 \pm 0.06,\  \delta=0.35 \pm 0.07.
\end{align}
In Fig.~\ref{Fig:shear_fit}, we compare $F(g^2T)$ and $f_\eta(g^2T)$ on the $32^3$ lattice. The fit function $F(x)$ describes $f_\eta(g^2T)$ in this $g^2T$ range well. These results imply that the shear viscosity of the CYM field is proportional to $1/g^{1.10-1.88}$ at weak coupling, which indicates that the shear viscosity in the classical limit depends more weakly on $g$ compared with the leading-order perturbative theory result, 
$\eta\propto 1/[g^4\ln{(g^{-1})}]$. Such a weak dependence on $g$ is consistent with the $g$-dependence of the "anomalous viscosity" under the strong disordered field, $\eta\propto 1/g^{1.5}$~\cite{ABM:shear}.

\subsection{Comparison with previous estimates}

We now compare our results with those of previous estimates obtained by analyzing the CYM field evolution in terms of the viscous hydrodynamic equation~\cite{EG:instability}. Since we have evaluated the shear viscosity using the Green-Kubo formula based on the linear response theory, the so-obtained shear viscosity characterizes the relaxation process around the equilibrium and is not sensitive to the initial conditions.

In Ref.~\cite{EG:instability}, Epelbaum and Gelis deduced the shear viscosity of the CYM field from the anisotropy of the pressure by comparing the energy density evolution in the expanding geometry to first-order viscous hydrodynamics. The shear viscosity in Ref.~\cite{EG:instability} was obtained
for a lattice spacing $a^{-1}=Q_s=2~{\rm GeV}$ and $g=0.5$ using the phenomenological glasma initial condition on a $64\times64\times128$ lattice.
The authors of Ref.~\cite{EG:instability} obtained the result $\eta/\varepsilon^{3/4}\simeq 0.3$, where $\varepsilon$ is the energy density.

\begin{figure}[htbp]
\begin{center}
\includegraphics[width=0.99\linewidth]{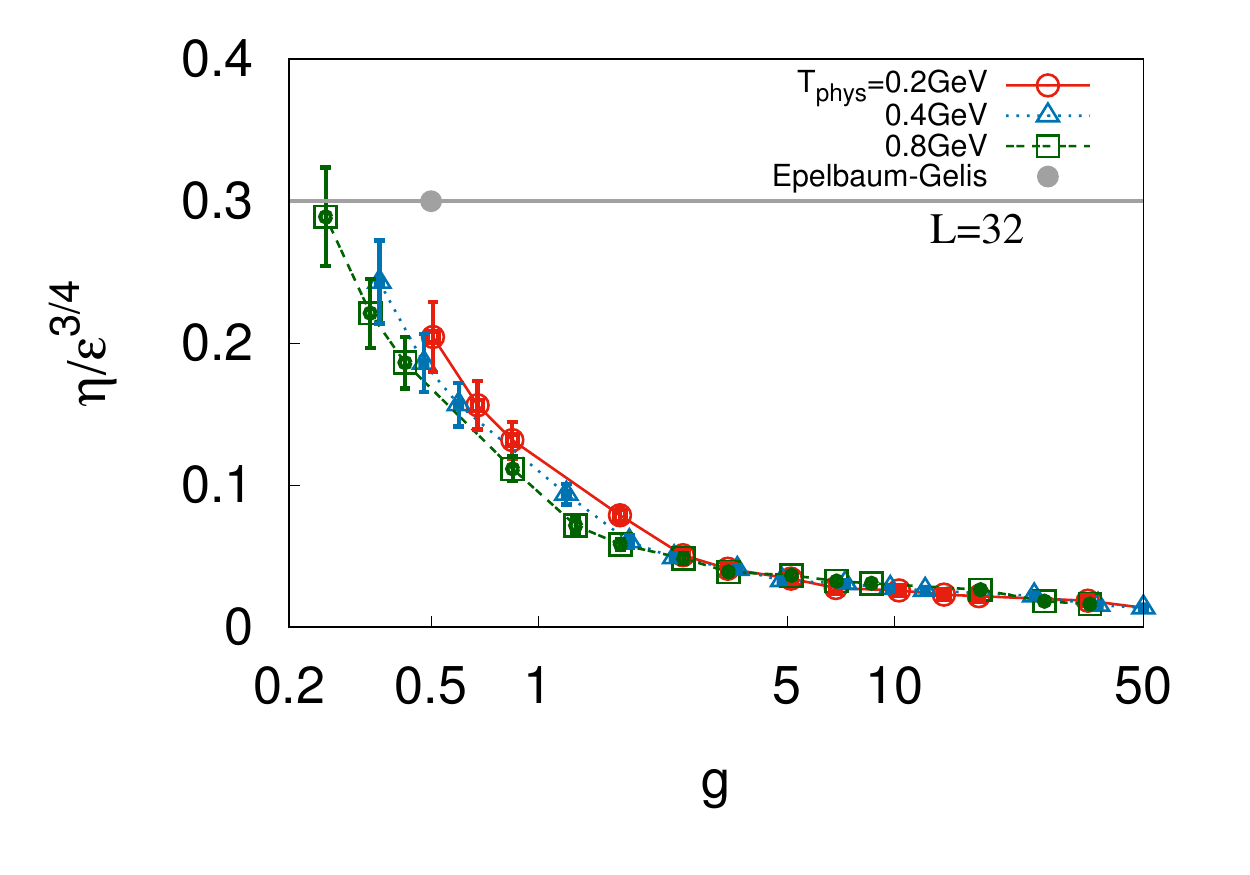}
\end{center}
\caption{
Shear viscosity normalized by $1/T$, $f_\eta=\eta/T$. The red, blue and green points show the results with $T_{\rm phys} = T/a = 0.2$~GeV, $0.4$~GeV and $0.8$~GeV, respectively, at $a=(2~{\rm GeV})^{-1}$. The black point shows the value of the shear viscosity in the expanding and anisotropic CYM field, which is estimated in~\cite{EG:instability}
}
\label{Fig:etaEG}
\end{figure}

In Fig.~\ref{Fig:etaEG}, we show the shear viscosity normalized by $\varepsilon^{3/4}$ with $\varepsilon \approx 3(N_c^2-1)T = 9T$ at lattice spacing $a = (2~{\rm GeV})^{-1}$ for the physical temperatures $T_\mathrm{phys}= T/a =0.2~{\rm GeV}, 0.4~{\rm GeV}$, and $0.8~{\rm GeV}$ as a function of $g$.
These temperatures in physical units are in the range of temperatures reached in relativistic heavy-ion collisions at RHIC and LHC at the start of the hydrodynamical evolution. The ratio $\eta/\varepsilon^{3/4}$ in the present work is smaller than the value obtained in Ref.~\cite{EG:instability} at $g=0.5$
by around 30-40 \%.
These differences may reflect the effect of a partial persistence of the initial condition and incomplete thermalization in the approach of Ref.~\cite{EG:instability}. However, we find it encouraging that our results obtained in equilibrium lie in the same numerical range as the shear viscosity value deduced from nonequilibrium dynamics in Ref.~\cite{EG:instability}.

It is worth noting that these values of the shear viscosity are much smaller than the leading-order perturbative result $\eta_\mathrm{pert}/T^3 \approx 1200$ for $g=0.5$~\cite{Arnold:shear}. This difference is conjectured~\cite{EG:instability} to be the manifestation of the anomalously small viscosity for systems made of strong disordered fields~\cite{ABM:shear}, an effect that is not included in the perturbative calculation.


\section{Summary}
\label{Sec:Summary}

We calculated the shear viscosity of the classical Yang-Mills (CYM) field on a lattice by applying the Green-Kubo formula to the time-correlation function $C(t)$ of the energy-momentum tensor in equilibrium. The time evolution of the CYM field was calculated for several values of the coupling constant $g$
starting from equilibrated configurations prepared at temperature $T \approx 1$, and $C(t)$ was evaluated as the ensemble average. The shear viscosity as a function of the coupling and temperature $\eta(g,T)$ was obtained from the low-frequency limit of the power spectrum of $C(t)$. The  dependence of the shear viscosity on $g$ and $T$ was represented as $\eta(g,T)/T=f_\eta(g^2T)$ with a scaling function $f_\eta$ in accordance with the scaling property of the CYM theory. The functional form of the scaling function  $f_\eta(g^2T)$ was extracted as a polynomial fit function.

The time-correlation function $C(t)$ was found to exhibit damped oscillatory behavior at early times followed by a slow decay. The damped oscillatory behavior is reflected in a broad bump around $\omega\approx 5$ in the power spectrum and may be caused by the decoherence among the two-momentum modes, which is attributed to a universal early-stage dynamics showing a damped-oscillatory behavior as observed in the case of the classical $\phi^4$ theory~\cite{Kyoto:scalarshear}.

The slowly decaying part of $C(t)$ produces a sharp peak at $\omega=0$ in the power spectrum. With decreasing $g^2T$, the height of the peak at $\omega=0$ increases and its width narrows. Thus the $\omega=0$ peak suggests the appearance of a collective mode with long life-time, which is most naturally identified with the hydrodynamic mode.

The scaling function $f_\eta$ rapidly increases with decreasing $g^2T$ at small $g^2T$ and its dependence on the coupling, $f_\eta \propto 1/g^{1.10-1.88}$, is found to be much weaker than the perturbative estimate. This weaker dependence may show the realization of anomalous viscosity
under the strong disordered field~\cite{ABM:shear}. The value of the shear viscosity in thermal equilibrium obtained in our analysis is found to be roughly consistent with that obtained from the glasma energy density evolution in the boost-invariant expanding geometry~\cite{EG:instability}. Thus the validity of the estimate of the shear viscosity from the energy density is confirmed.

While the transport properties of the CYM field around equilibrium may be relevant to the small shear viscosity in high-energy heavy-ion collisions, there still remain problems to be considered in order to discuss the shear viscosity around equilibrium in a more rigorous way. The behavior of high-dimensional observables such as the time-correlation function of the energy-momentum tensor, $C(t)$, is dominated by hard thermal particles with momenta of order temperature and is sensitive to the ultraviolet cutoff in the classical field theory. In addition, the equipartition property of classical fields leads to the Rayleigh-Jeans divergence in the continuum limit, $a \to 0$.
It should be remembered, however, that the scaling function allows us to evaluate the shear viscosity in the low momentum region by using the classical Yang-Mills field simulation at large lattice temperatures, where the classical field description is justified.

There are several sophisticated ways to circumvent these problems
on the contribution of high-momentum contribution, such as
renormalizing the theory by introducing counterterms in the classical field~\cite{Renormalization},
applying the effective dynamics obtained by integrating hard particles~\cite{HardParticle},
taking account of the explicit coupling of fields and particles~\cite{Field_Particle},
and
switching to dynamics described by the kinetic theory~\cite{Kurkela}.
Applying the classical field theory with quantum statistical nature~\cite{Replica} would be another interesting direction to be studied.
%

\section*{Acknowledgments}
The authors would like to thank Michael Strickland and Gert Aarts
for useful comments.
This work is supported in part by the Grants-in-Aid for Scientific Research
from JSPS (Nos. 
19K03872, 
19H01898, 
and
19H05151), 
the Yukawa International Program for Quark-hadron Sciences (YIPQS),
and the U.S.~Department of Energy (grant DE-FG02-05ER41367).

\appendix


\section{Local electric density distribution}\label{App:local}

We here discuss the equilibration of the configurations used as an ensemble by checking the distribution of the local electric density
defined as
\begin{align}
\varepsilon_E(x)=\frac12\sum_{a} E^a_1(x)^2.
\end{align}
If the configurations reach equilibrium, the distribution of $\varepsilon_E(x)$ in the configurations is given by the following functional form,
\begin{align}
P[\varepsilon_E(x)]\propto\sqrt{\varepsilon_E(x)}
\exp[-\varepsilon_E(x)/T].
\label{Eq:epsEdist}
\end{align}
This thermal distribution function can be obtained from the canonical partition function in Eq.~\eqref{partition},
$(\prod_a dE^a_1)\,\exp(-\varepsilon_E/T)=\sqrt{2\varepsilon_E}d\varepsilon_E d\Omega\,\exp(-\varepsilon_E/T)$, where $\Omega$ is the solid angle in the 3 dimensional color space. 

\begin{figure}[htbp]
\begin{center}
\includegraphics[width=0.99\linewidth]{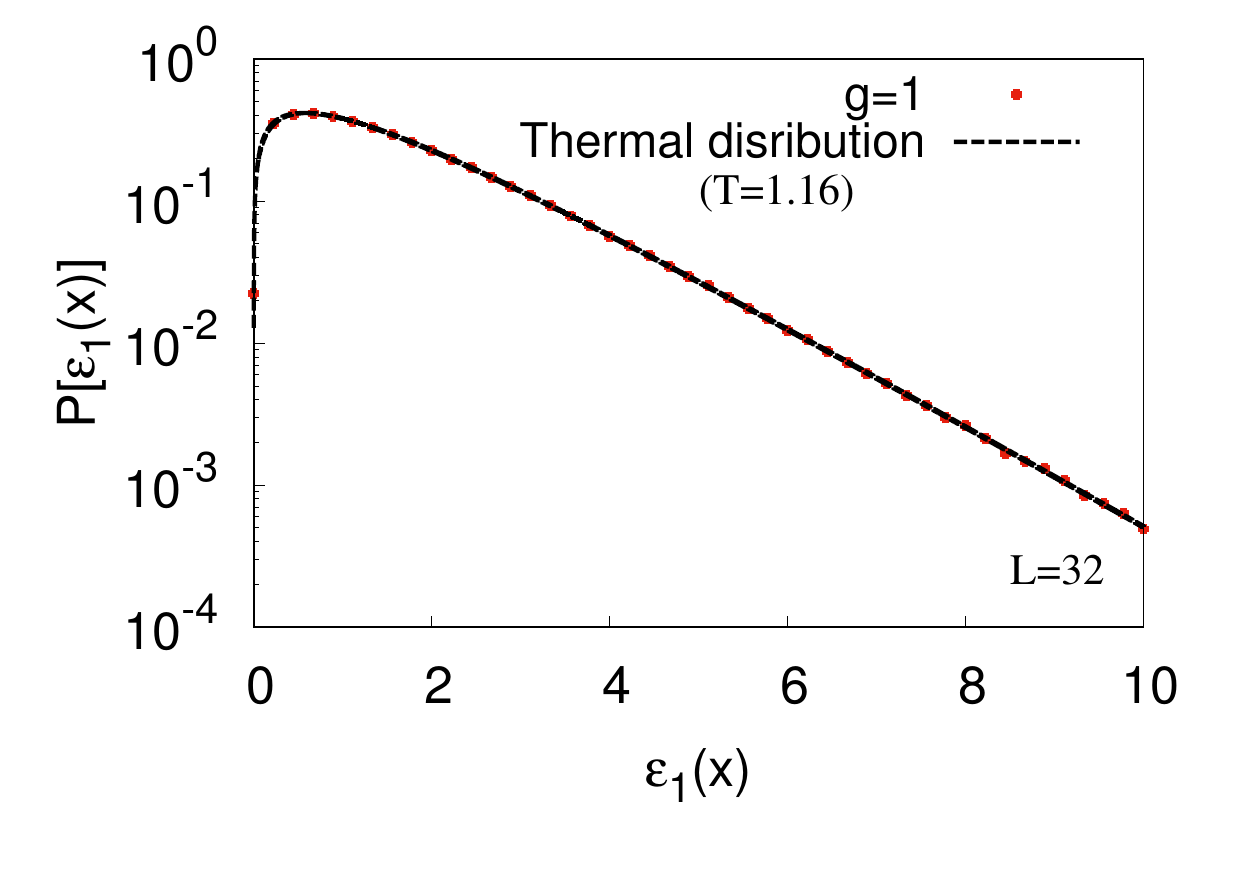}
\end{center}
\caption{Distribution of $\varepsilon_E(x)$ in the configurations at $t-t_{\rm ini}=200$ with $g=1$ on the $32^3$ lattice.
The black dashed line show the thermal distribution function with $T=1.16$($g^2T=1.16$) 
in Eq.~\eqref{Eq:epsEdist}
}
\label{Fig:loc}
\end{figure}

In Fig.~\ref{Fig:loc}, we show the distribution of $\varepsilon_E(x)$ in the configurations at $t-t_{\rm ini}=200$ with $g=1$ on the $32^3$ lattice.
This numerical distribution agrees with the thermal distribution whose temperature is $1.16$($g^2T=1.16$).
From this agreement, we conclude that the configurations
are sufficiently thermalized at $t-t_{\rm ini}=200$.

\section{Power spectrum for $g^2T \gg 1$}\label{App:strong}

\begin{figure}[tbhp]
\begin{center}
\includegraphics[width=0.99\linewidth]{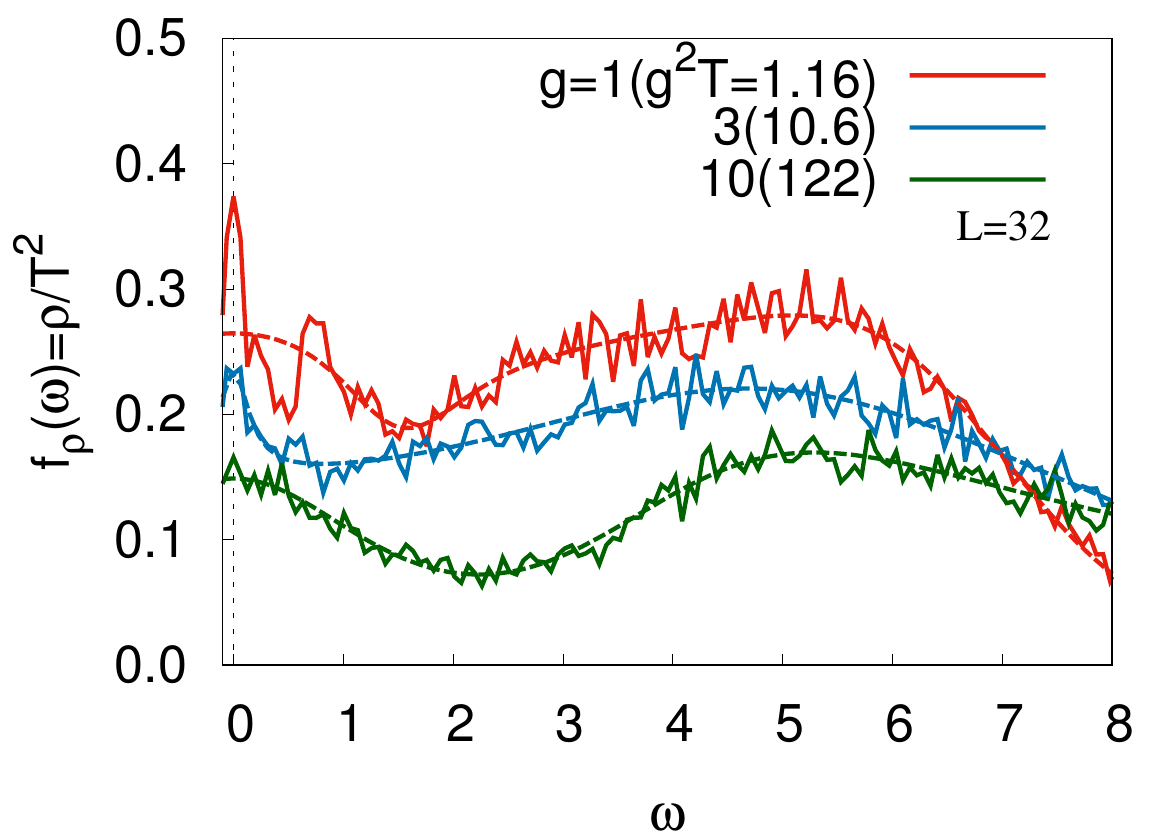}
\end{center}
\caption{Fourier spectra normalized by $T^2$, $f_\rho(\omega)=\rho(\omega)/T^2$, at $g=1, 3$ and $10$($g^2T=1.16, 10.6$ and $122$) on the $32^3$ lattice by solid curves.
The dashed curves depict the fitting function given as the Fourier transform of the exponential decay plus damped oscillations in Eq.~\eqref{Eq:expDO}.
}
\label{Fig:Spec2}
\end{figure}

The power spectra in the very large $g^2T$ region also show interesting features. In Fig.~\ref{Fig:Spec2}, we show the normalized power spectrum, $f_\rho(\omega)=\rho(\omega)/T^2$ at $g=1, 3, 10$ ($g^2T=1.16, 10.6, 122$) by solid curves. In this $g^2T \gg 1$ region, $f_\rho(\omega)$ consists of
the peak at $\omega=0$ and the bump at $\omega=(5-7)$. 

This structure is well described by the Fourier transform of the exponential decay and damped oscillations,
\begin{align}
F_{{\rm exp}+{\rm DO}}(t)=a_0e^{-b_0t}+\sum_{i=1,2} a_ie^{-b_it}\cos{(c_it+d_i)}.
\label{Eq:expDO}
\end{align}

Dashed curves show the Fourier transform of $F_{{\rm exp}+{\rm DO}}(t)$, which almost completely agrees with the numerical results and thus is hard to discern in the figure. In particular, the result obtained at the strongest coupling, $g=10$ ($g^2T=116$), agrees well with the fit. This agreement of the functional form may be due to the dominance of the four-point interaction both in the large $g^2T$ region of the CYM theory and in the classical massless $\phi^4$ theory.

\clearpage

\end{document}